\documentclass[aps,prl,reprint,superscriptaddress,showpacs,amsmath,amssymb,longbibliography,lengthcheck]{revtex4-1}
\usepackage{graphicx}
\usepackage{mathrsfs}
\usepackage{txfonts}

\begin{document}

 \title{Damping of Quantum Vibrations Revealed in Deep Sub-barrier Fusion} 

\author{Takatoshi Ichikawa}%
\affiliation{Yukawa Institute for Theoretical Physics, Kyoto University,
Kyoto 606-8502, Japan}

\author{Kenichi Matsuyanagi}
\affiliation{Yukawa Institute for Theoretical Physics, Kyoto University,
Kyoto 606-8502, Japan}
\affiliation{RIKEN Nishina Center, Wako 351-0198, Japan}
\date{\today}

\begin{abstract}
 We demonstrate that when two colliding nuclei approach each other,
 their quantum vibrations are damped near the touching point. We show
 that this damping is responsible for the fusion hindrance phenomena
 measured in the deep sub-barrier fusion reactions.  To show those, we
 for the first time apply the random-phase-approximation (RPA) method to
 the two-body $^{16}$O + $^{16}$O and $^{40}$Ca + $^{40}$Ca systems. We
 calculate the octupole transition strengths for the two nuclei
 adiabatically approaching each other. The calculated transition
 strength drastically decreases near the touching point, strongly
 suggesting the vanishing of the quantum couplings between the relative
 motion and the vibrational intrinsic degrees of freedom of each
 nucleus. Based on this picture, we also calculate the fusion cross
 section for the $^{40}$Ca + $^{40}$Ca system using the coupled-channel
 method with the damping factor simulating the vanishing of the
 couplings. The calculated results reproduce well the experimental data,
 indicating that the smooth transition from the sudden to adiabatic
 processes indeed occurs in the deep sub-barrier fusion reactions.
\end{abstract}


\maketitle

Heavy-ion fusion reactions at low incident energies serve as an important
probe for investigating the fundamental properties of the potential
tunneling of many-body quantum systems.  When two nuclei fuse, a
potential barrier, called the Coulomb barrier, is formed because of the
strong cancellations between the Coulomb repulsion and the attractive
nuclear force. The potential tunneling at incident energies below this
Coulomb barrier is called the sub-barrier fusion.  One important aspect
of the sub-barrier fusion reactions is couplings between the relative
motion of the colliding nuclei and nuclear intrinsic degrees of
freedom, such as collective vibrations of the target and/or projectile
\cite{DHRS98}. Those couplings result in the large enhancement of the
fusion cross sections at the sub-barrier incident energies, as compared
to the estimation of a simple potential model.  The coupled-channel (CC)
model taking into account the couplings has been successful in
explaining this enhancement \cite{Bal98,Hagino12}.

Recently, it has been possible to measure the fusion cross sections down
to extremely deep sub-barrier incident energies
\cite{Jiang02,das07,stef08,Mont12}. The unexpected steep fall-off of the
fusion cross sections, compared to the standard CC calculations, emerges
at the deep sub-barrier incident energies in a wide range of mass
systems.  These steep fall-off phenomena are often called the fusion
hindrance. The emergence of the fusion hindrance shows the threshold
behavior, which is strongly correlated with the energy at the touching
point of the two colliding nuclei \cite{ich07-2,EIS12}. In this respect, it
has been shown that a key point to understand this fusion hindrance is
the potential tunneling in the density overlap region of the two
colliding nuclei (see Fig.~1 in Ref.~\cite{ich07-2}).

To describe the fusion hindrance phenomena, many theoretical models to
extend the standard CC model have been proposed. Based on the sudden
picture, Mi\c{s}icu and Esbensen have proposed that a strong repulsive
core exists in the inner part of the Coulomb barrier due to nuclear
incompressibility \cite{mis06}. This model can reproduce well the fusion
hindrance from the light- to heavy-mass systems
\cite{mis06,Esb10,mis11,Mont12}. Dasgupta {\it et al.}~proposed the
concept of the quantum decoherence of the channel couplings
\cite{das07}, but there are only simple calculations with this model
\cite{dia08}. Based on the adiabatic picture, which is the opposite
limit to the sudden approach, Ichikawa {\it et al.}~introduced the
damping factor in the standard CC calculations in order to smoothly
joint between the sudden and adiabatic processes \cite{ich09}. This
model can reproduce the fusion hindrance better than the sudden
model. However, the physical origin of the damping factor was still
unclear.

In this Letter, we show the physical origin of the damping factor
proposed in Ref.~\cite{ich09}.  In the standard CC model, it has been
assumed that the properties of the vibrational modes do not change, even
when two colliding nuclei touch with each other. However, in fact, the
single-particle wave functions drastically change in the two nuclei
approaching each other. This results in the damping of the vibrational
excitations, that is, the vanishing of the couplings between the
relative motion and the vibrational excitations of each nucleus.  To
show this, we for the first time apply the random-phase approximation
(RPA) method to the two-body $^{16}$O + $^{16}$O and $^{40}$Ca +
$^{40}$Ca systems and calculate the octupole transition strength,
$B(E$3), as a function of the distance between the two nuclei. We below
show that the obtained $B(E$3) values for the individual nuclei are
indeed damped near the touching point.

To illustrate our main idea, we first discuss a disappearance of the
octupole vibration during the $^{16}$O + $^{16}$O fusion process.  We
calculate the mean-field potential with the folding procedure using the
single Yukawa function to conserve its inner volume~\cite{Bol72}. In the
two-body system before the touching point, we assume the two
sharp-surface spherical nuclei. After the touching point, we describe
the nuclear shapes with the lemniscatoids parametrization, as shown in
Ref.~\cite{ich07-1}.  Using this, we can describe the smooth transition
from the two- to one-body mean-field potentials.  The depths of the
neutron and proton potentials are taken from Ref.~\cite{Mol95}. We use
the radius for the proton and neutron potentials, $R_0$, with $R_0=1.26
A^{1/3}$ fm, where $A$ is the total nucleon number. In the calculations,
the origin is located at the center-of-mass position of the two nuclei.
The above procedure works well, because Umar and Oberacker found that
the double-folding potential with the frozen density agrees almost
perfectly with the density-constrained time-dependent Hartree-Fock
(TDHF) approach for distances $R \ge 6$ fm \cite{Umar06}.

\begin{figure}[t]
\includegraphics[keepaspectratio,scale=0.65]{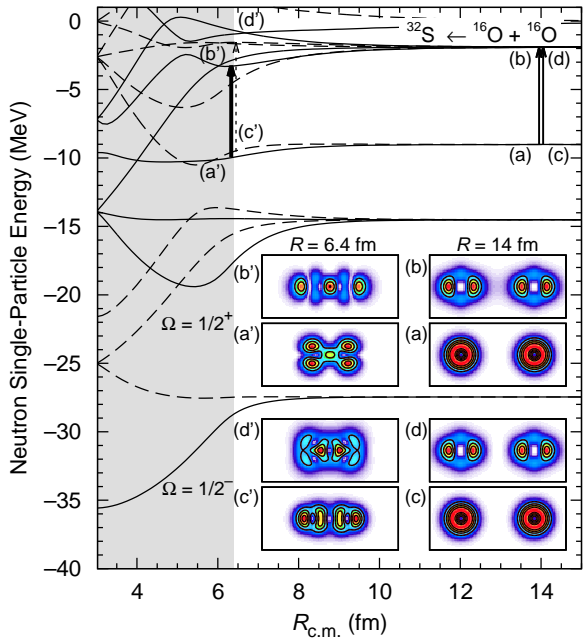}
\caption{(color online) Nilsson diagram for the neutron single-particle
states versus the distance between $^{16}$O + $^{16}$O. The solid and
dashed lines denote the positive and negative party states,
respectively. The gray area denotes the overlap region of $^{16}$O +
$^{16}$O. The solid and dashed arrows denote the main $p$-$h$
excitations generating the octupole vibration. The single-particle
density distributions for these $p$ and $h$ states are given in the
insets from (a) to (d) and (a') to (d').}
\end{figure}

Using the obtained mean-field potentials, we solve the axially-symmetric
Schr\"odinger equation with the spin-orbit force.  Then, the parity,
$\pi$, and the $z$ component of the total angular momentum, $\Omega$,
are the good quantum numbers.  The details of the model and the
parameters are similar to Refs.~\cite{Bol72,Mol95}.  We calculate the
single-particle wave functions of both the projectile and target using
the one-center single Slater determinant. We expand the total
single-particle wave function by many deformed harmonic-oscillator bases
in the cylindrical coordinate representation.  The deformation parameter
of the basis functions is determined so as to cover the target and
projectile. The basis functions are taken with its energy lower than 25
$\hbar\omega$.

\begin{figure}[t]
\includegraphics[keepaspectratio,scale=0.6]{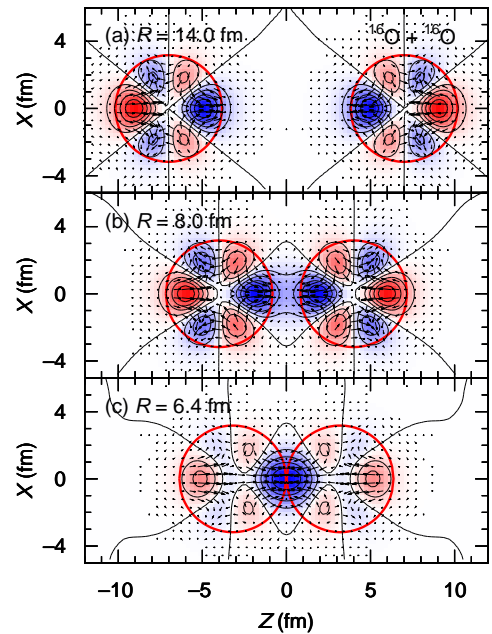}
\caption{(color
online) Transition densities and currents for the first excited $3^-$
state with $\Omega^\pi=0^+$ at $R=$ (a) 14.0 fm, (b) 8.0 fm, and (c) 6.4
fm.  The contour lines denote the transition density. The arrows denote
the current density. These two values are normalized in each plot.  The
(red) thick solid line denotes the half depth of the mean-field
potential.}
\end{figure}

Figure 1 shows the Nilsson diagram for the obtained neutron
single-particle energies versus the distance between $^{16}$O +
$^{16}$O.  The solid and dashed lines denote the positive and negative
party states, respectively.  The gray area denotes the overlap region of
the two nuclei. The distance $R=6.4$ fm corresponds to the touching
point.  Some densities for the obtained single particles at $R=14$
and 6.4 fm are given in the insets from (a) to (d) and (a') to (d') in
Fig.~1, respectively.  At $R=14$ fm, we see that the positive-
(negative-) parity indicates the symmetric (asymmetric) combinations of
the single-particle states referring to the right- and left-sided
$^{16}$O. Thus, the positive- and negative-parity single-particle states
are degenerate for large $R$.  With decreasing $R$, these
single-particle states smoothly change to those for the composite
$^{32}$S system.

We can now easily extend the RPA method \cite{Ring} to the two-body
system, because the wave functions of both the one- and two-body systems
are described with the one-center Slater determinant. We can directly
superpose all combinations of the particle ($p$) and hole ($h$) states
for the obtained single particles in a unified manner for both the one-
and two-body systems. We solve the RPA equation at each center-of-mass
distance between $^{16}$O + $^{16}$O.  At large $R$ values, the RPA
solutions with $\Omega^\pi=0^+$ and $0^-$ represent the symmetric and
asymmetric combinations of the states where the RPA modes are excited in
either the right- or left-sided $^{16}$O.  When $R$ decreases below the
touching point, they smoothly change to excitation modes in the
composite $^{32}$S system.  In the calculations, we only take into
account the $p$-$h$ states with the excitation energies below 30 MeV.
We use the residual interaction as the density-dependent contact one
taken from Ref.~\cite{SB75}.  The strength of the residual interaction
is determined at each $R$ such that the lowest $\Omega^{\pi}=0^-$
solution of the RPA appears at zero energy.  We have developed a new
scheme based on the Tomonaga theory of collective motion
\cite{Tomo55,ich99}, that enables us to separate the center of mass, the
relative motion, and the intrinsic degrees of freedom. It is a
generalization of the known procedure in the RPA \cite{Kam98}.
It is also interesting to compare our results with the TDHF dynamics
\cite{Keser12}. 

\begin{figure}[t]
\includegraphics[keepaspectratio,scale=0.58]{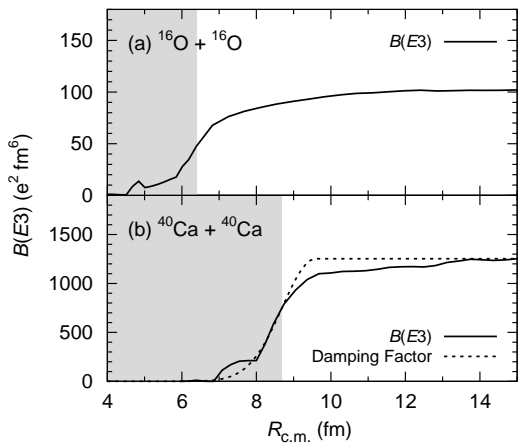}\\%
\caption{Transition strengths, $B$($E$3), for the first excited state
versus the distance between (a) $^{16}$O + $^{16}$O and (b) $^{40}$Ca +
$^{40}$Ca.  The solid line denotes the calculated results. The gray area
denotes the overlap region of the two colliding nuclei.
In the bottom panel (b), the
dotted line denotes the damping factor of Eq.~(2) in Ref.~\cite{ich09}
estimated from the experimental data of the fusion cross section.}
\end{figure}


For the calculated results, the obtained first-excited state of the
right-sided $^{16}$O is the octupole ($3^-$) one with a large $B(E$3)
value. At $R=15$ fm, the excitation energy and the $B(E3,
3_1^-\to0_1^+)$ value are 5.29 MeV and 102.07 e$^2$ fm$^6$,
respectively. We have checked that those values are consistent with the
calculated results of the one-body $^{16}$O.  Figure 2(a) shows the
calculated transition densities and currents \cite{Ring} for the first
excited state with $\Omega^\pi=0^+$ at $R=14$ fm. In Fig. 2(a), we can
clearly see the octupole vibrations in both $^{16}$O's.  At $R=8.0$ fm,
the transition density of the neck part between two $^{16}$O's develops
[see Fig. 2(b)]. At $R=6.4$ fm, the octupole vibrations of each $^{16}$O
become weak [see Fig. 2(c)].
The degenerating excitation energies of the first-excited states with
$\Omega^\pi=0^+$ and $0^-$ split below $R=8$ fm.
They become 5.82 and 4.83 MeV at $R=6.4$ fm.

To more clearly see the damping of the octupole vibrations, we calculate
the $B(E$3) value for the right-sided $^{16}$O.
 We can easily calculate it by taking a symmetric
linear combination of the octupole transition amplitudes for the
positive- and negative-parity RPA solutions.
Figure 3 (a) shows the
calculated $B(E3, 3_1^-\to0_1^+)$ values versus the distance between
$^{16}$O + $^{16}$O. 
In Fig.~3 (a), we can see that the $B(E3)$ value for the right-sided
$^{16}$O (the solid line) falls off at
around $R=8$ fm with decreasing $R$, indicating that the octupole
vibrations are strongly suppressed near the touching point.

The damping of the vibrations originates from the change of the
single-particle wave functions. At $R = 14$ fm, the major $p$-$h$
excitations generating the octupole vibration are those from the
$p_{1/2}$ to the $d_{5/2}$ single-particle states in $^{16}$O.
In these states, the degenerating positive- and negative-parity doublet
states contribute equally to generate the octupole vibration.  Those can
be seen in the density distributions of the $\Omega^{\pi}=1/2^+$ and
$1/2^-$ states given in the insets from (a) to (d) in Fig.~1.  The
corresponding $p$-$h$ excitations are denoted by the solid arrows from
(a) to (b) and (c) to (d) in Fig.~1. When the two nuclei approach each
other, the features of these single-particle wave functions drastically
change.  At $R = 6.4$ fm, the neck formation takes place in the
positive-parity states, while it is forbidden for the negative-parity
states (having nodes at the touching point). Thus, the density
distributions of those parity partners become quite different from each
other. [see the insets from (a') to (d')].  In the RPA calculation, the
contributions from the negative-parity states to the octupole vibration
become small at the touching point [see the solid and dotted arrow from
(a') to (b') and (c') to (d'), respectively], resulting in the decreases
of the $B(E3)$ value.

The mechanism for the damping of the quantum vibration is a general one
valid also for heavier mass systems.  We have also performed the RPA
calculations for the $^{40}$Ca + $^{40}$Ca and $^{56}$Ni + $^{56}$Ni
systems. In the calculations, we use $R_0=1.27A^{1/3}$. We obtained the
similar damping of the $B(E$3) values for the both systems. In
Fig.~3(b), the solid line denotes the calculated result for the
$^{40}$Ca + $^{40}$Ca system.  The calculated excitation energy and
$B(E3, 3_1^-\to0_1^+)$ value for the $^{40}$Ca +$^{40}$Ca system are
3.24 MeV and 1253.17 $e^2$ fm$^6$ at $R=15.0$ fm, respectively.

As shown above, the octupole vibrations are damped near the touching
point, resulting in the vanishing of the couplings between the relative
motion and the vibrational intrinsic degrees of freedoms.  This
vanishing would lead to the smooth transition from the sudden to
adiabatic process, as shown in Ref.~\cite{ich09}. It is clear that such
effect has not been taken into account in the standard CC model.  One
candidate to include it is the introducing of the damping factor
proposed in Ref.~\cite{ich09}.  To clearly see the effect of the
vanishing of the couplings, we calculate the fusion cross section for
the $^{40}$Ca + $^{40}$Ca system using the computer code {\tt
CCFULL}~\cite{ccfull} coupled with the damping factor based on the model
of Ref.~\cite{ich09}.

In the calculations, we include the couplings to only the low-lying
$3^-$ and $2^+$ states and to single phonon and all mutual excitations
of these states.  We take the energies and the deformations of each
state taken from Ref.~\cite{Mont12} to reproduce well the experimental
data. We use the same deformation 
parameters for the Coulomb and nuclear couplings.  For the parameters of
the Yukawa-plus-exponential (YPE) model, we use $r_0=1.191$ fm and
$a=0.68$ fm.

It is remarkable that the damping factor strongly correlates with the
calculated $B(E$3) value for the $^{40}$Ca + $^{40}$Ca system.  To show
this, we take $\lambda_\alpha=0$ in the damping factor of Eq.~(2) in
Ref.~\cite{ich09} for simplicity.  We tune the parameters $R_d$ and
$a_d$ in the damping factor so as to reproduce the experimental data of
the fusion cross section. We obtain $R_d=9.6$ fm and $a_d=0.9$ fm as the
best fit to the data.  In Fig 3(b), the dotted line denotes the obtained
damping factor normalized at $R=15$ fm.

\begin{figure}[t]
\includegraphics[keepaspectratio,scale=0.7]{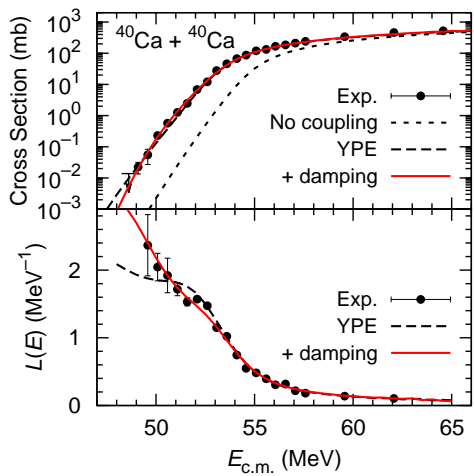}\\%
\caption{(color
online) Fusion cross sections (upper panel) and its logarithmic
derivative (lower panel) for the $^{40}$Ca + $^{40}$Ca system versus the
incident energies.  The solid circles denote the experimental data taken
from Ref.~\cite{Mont12}.  The solid and dashed lines denote the
calculated results of the coupled-channel method using the YPE potential
with and without the damping factor, respectively.  The dotted line
denotes the calculated result without the couplings.  }
\end{figure}

Figure 4 shows the calculated fusion cross section (upper panel) and its
logarithmic derivative $d\ln(E_{\rm c.m.}\sigma_{\rm fus})/dE_{\rm
c.m.}$ (lower panel) for the $^{40}$Ca + $^{40}$Ca system. The solid and
dashed lines denote the calculation with and without the damping factor,
respectively. The dotted line denotes the calculation without the
couplings. In Fig.~4, we can see that the calculated results with the
damping factor reproduce well the experimental data, which is better
than the sudden model \cite{mis11,Mont12}.  In our model, the calculated
astrophysical S-factor has a peak structure.  We also performed the CC
calculation for the $^{48}$Ca + $^{48}$Ca system and the calculated
result reproduces well the experimental data.  The CC calculations with
the damping factor also already reproduced well the experimental data
for the $^{64}$Ni + $^{64}$Ni, $^{58}$Ni + $^{58}$Ni, and $^{16}$O +
$^{208}$Pb systems \cite{ich09}.

In summary, we have demonstrated the damping of the quantum vibrations
when two nuclei adiabatically approach each other.  To show this, we
for the first time applied the RPA method to the two-body $^{16}$O +
$^{16}$O and $^{40}$Ca + $^{40}$Ca systems and calculated the $B(E$3)
values of each nucleus. We have shown that the calculated $B(E3$)
value is indeed damped near the touching point. We have also shown that
the damping factor proposed in Ref.~\cite{ich09} strongly correlates
with the calculated $B(E$3) values and the calculations of the fusion
cross section coupled with the damping factor reproduce well the
experimental data. This indicates that the fusion hindrance originates
from the damping of the quantum couplings and strongly suggests that the
smooth transition from the sudden to adiabatic processes occurs near
the touching point.

 \begin{acknowledgments}
  The authors thank K.~Hagino and A.~Iwamoto for useful discussions.  A
  part of this research has been funded by MEXT HPCI STRATEGIC PROGRAM.
  This work was undertaken as part by the Yukawa International Project
  for Quark-Hadron Sciences (YIPQS).
 \end{acknowledgments}

\end{document}